\documentclass[twocolumn]{aastex701}
\usepackage{amsmath}
\usepackage{bm}
\usepackage{subfigure}
\usepackage{placeins}
\usepackage{xcolor}

\begin{document}

\title{Direct Numerical Simulations of Oxygen-Flame-Driven Deflagration-to-Detonation Transition in Type \uppercase\expandafter{\romannumeral1}a Supernovae}

\correspondingauthor{Lile Wang and Yao Zhou}
\email{lilew@pku.edu.cn, yao.zhou@sjtu.edu.cn}

\author[orcid=0009-0006-4343-3465]{Xiaoyu Zhang}
\affiliation{School of Physics and Astronomy, Institute of Natural Sciences and MOE-LSC, Shanghai Jiao Tong University, Shanghai 200240, China}
\email{xiaoxiangyeyu@sjtu.edu.cn}  

\author[0000-0002-6540-7042]{Lile Wang}
\affiliation{The Kavli Institute for Astronomy and Astrophysics, Peking University, Beijing 100871, China}
\affiliation{Department of Astronomy, School of Physics,
  Peking University, Beijing 100871, China} 
\email{lilew@pku.edu.cn}

\author[0000-0002-6316-1632]{Yang Gao}
\affiliation{School of Physics and Astronomy, Sun Yat-Sen University, Zhuhai, Guangdong 519082, China}
\email{gaoyang25@mail.sysu.edu.cn}

\author[0000-0002-3616-2912]{Yao Zhou}
\affiliation{School of Physics and Astronomy, Institute of Natural Sciences and MOE-LSC, Shanghai Jiao Tong University, Shanghai 200240, China}
\email{yao.zhou@sjtu.edu.cn}

\begin{abstract}
We present direct numerical simulations demonstrating deflagration-to-detonation
transition (DDT) driven by oxygen flames in Type Ia supernova progenitors.
Using the Castro hydrodynamics code coupled with the ``aprox13''
13-isotope nuclear network, we simulate combustion in isolated fuel regions where
oxygen flames trail carbon flames. In a fiducial one-dimensional run at
$\rho_{0}=3.5\times10^{7}\ \mathrm{g\ cm^{-3}}$ we observe spontaneous DDT of the
oxygen flame via the Zel'dovich gradient mechanism when the carbon–oxygen separation
reaches $\sim 10\ \mathrm{km}$. The oxygen detonation then captures the carbon
flame and triggers a stable carbon detonation. Systematic one-dimensional parameter scans show
that successful carbon DDT requires upstream densities in the range
$(3.1$--$3.6)\times10^{7}\ \mathrm{g\ cm^{-3}}$ and a minimum carbon-flame
thickness of $\gtrsim 20\ \mathrm{m}$. Two-dimensional simulations confirm DDT
and demonstrate that the multidimensional cellular structure of the oxygen
detonation can promote carbon detonation
at somewhat lower densities than in one dimension. These results provide direct numerical
evidence that oxygen-flame-driven DDT is physically plausible in turbulent
white-dwarf environments and underscore the importance of multidimensional
effects for Type Ia supernova explosion modeling.
\end{abstract}


\section{Introduction} \label{sec.introduction}

Type Ia supernovae (SNe~\uppercase\expandafter{\romannumeral1}a) serve as reliable standard candles and play a pivotal role as distance indicators in constructing the cosmic distance ladder \citep{kirshner2009foundationssupernovacosmology}, owing to their remarkably uniform and standardizable light curves. Despite their critical importance in cosmology, the exact nature of their progenitor systems and explosion mechanisms remains one of the central open problems in stellar astrophysics. This uncertainty largely arises from the lack of direct observational evidence that can probe the thermonuclear runaway and flame propagation inside white dwarfs.

The majority of observed SNe~Ia, commonly referred to as ``normal'' events \citep{branch_relative_1993}, are thought to originate from the thermonuclear disruption of a carbon--oxygen white dwarf close to the Chandrasekhar mass limit. For such a progenitor, the central density can reach $2$--$6\times10^{9}\ \rm{g\ cm^{-3}}$ \citep{PhysRevC.72.025806}, leading to spontaneous ignition and the onset of a thermonuclear deflagration (subsonic burning). During the early deflagration phase, the star expands, decreasing the density in the outer layers. When the fuel density drops to the regime where burning primarily synthesizes intermediate-mass elements, the flame can transition into a supersonic detonation that unbinds the entire star. This transition from a subsonic to a supersonic combustion mode is known as the deflagration-to-detonation transition (DDT) \citep{gamezo_flame_2008}, and is widely believed to be a key element in explaining the observed nucleosynthetic yields and light curves of SNe~Ia.

In the absence of external shocks, the Zel’dovich gradient mechanism \citep{zeldovich_onset_1970} provides a natural pathway for DDT. If a sufficiently steep temperature gradient exists within the unburned fuel, spontaneous energy release can couple to a developing shock, resulting in mutual amplification and ultimately a self-sustained detonation wave. Extensive numerical studies have investigated this mechanism in the context of carbon flames. For example, \citet{khokhlov_deflagrationdetonation_1997} reported the occurrence of DDT in direct simulations, identifying a favorable density range of $0.5$--$5\times10^{7}\ \rm{g\ cm^{-3}}$. Later, \citet{woosley_type_2009} examined how turbulence interacts with carbon flames and argued that successful DDT requires Damköhler numbers (a dimensionless parameter that compares the characteristic reaction time scale to the transport time scale) of order unity to ten, turbulent velocities exceeding $500\ \rm{km\ s^{-1}}$, and ambient fuel densities around $10^{7}\ \rm{g\ cm^{-3}}$. More recently, \citet{gusto_toward_2024} employed three-dimensional turbulent simulations to map out a more realistic parameter space for detonation initiation.

While most previous work has focused on DDT in carbon flames, \citet{woosley_flames_2011} proposed that oxygen flames could also trigger detonations under certain conditions. At fuel densities of $2$--$3\times10^{7}\ \rm{g\ cm^{-3}}$, carbon burning proceeds much more rapidly than oxygen burning, leaving behind an extended region of partially burned material in which oxygen remains unburned. If turbulence broadens this separation to $\sim10\ \rm{km}$, the oxygen flame itself may undergo DDT through the Zel’dovich gradient mechanism. Building on this idea, \citet{shu_flame_2020} analytically demonstrated that vortices generated by Rayleigh--Taylor (R--T) instabilities can, through turbulent mixing and flame stretching, generate oxygen-rich regions of sufficient spatial extent to undergo DDT. However, despite these theoretical developments, direct numerical confirmation of oxygen-flame-driven DDT has been lacking.

In this work, we address this issue by performing high-resolution hydrodynamic simulations to directly examine DDT triggered by oxygen flames. We assume that a sufficiently wide separation has developed between the carbon and oxygen flame fronts, as might be established by turbulence in a realistic white dwarf interior. Under this condition, we first carry out one-dimensional (1D) direct numerical simulations to determine the density and carbon flame thickness required for successful DDT. We find that carbon DDT can occur only within a fuel density range of $3.1$--$3.6\times10^{7}\ \rm{g\ cm^{-3}}$ and when the carbon flame thickness exceeds $\sim 20\ \rm{m}$. We then extend our study to two-dimensional (2D) simulations, in which multidimensional detonation structures---including triple points and transverse waves---naturally emerge. These structures facilitate carbon detonation even at densities lower than those required in 1D, demonstrating the crucial role of multidimensional effects in realistic astrophysical environments.

The remainder of this paper is organized as follows. Section~\ref{sec.formulation} describes the physical background and numerical methods used in this study. Section~\ref{1D results} presents the setup and results of 1D simulations, focusing on the conditions for oxygen-initiated DDT. Section~\ref{2D results} discusses the 2D simulations and the role of multidimensional structures in enabling carbon detonation at lower densities. Finally, Section~\ref{summary} summarizes our main findings and outlines future research directions.

\section{Formulation of the Problem} \label{sec.formulation}
\subsection{Background}

\begin{figure}[htbp]
\centering
\subfigure[]{
\label{Fig.global_island}
\includegraphics[width=0.45\textwidth]{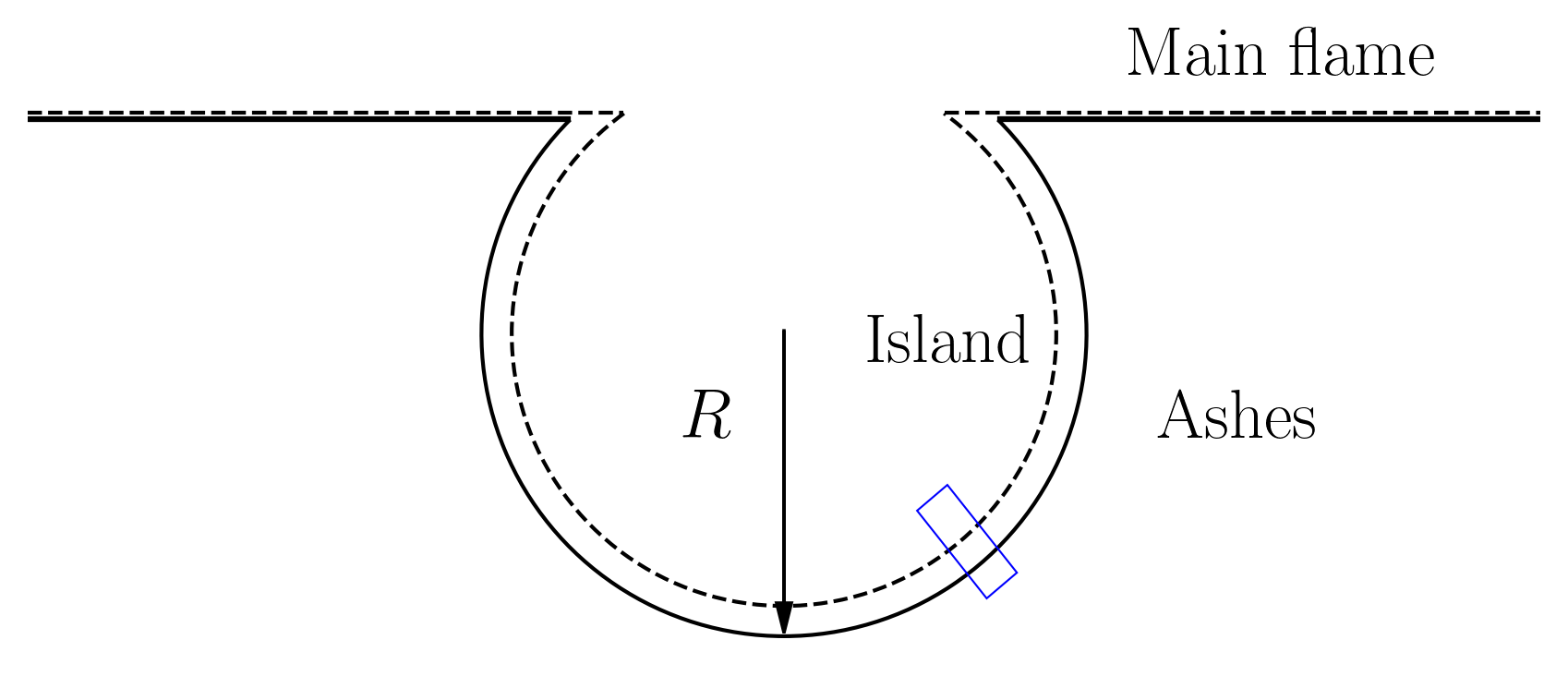}}
\subfigure[]{
\label{Fig.local_island}
\includegraphics[width=0.45\textwidth]{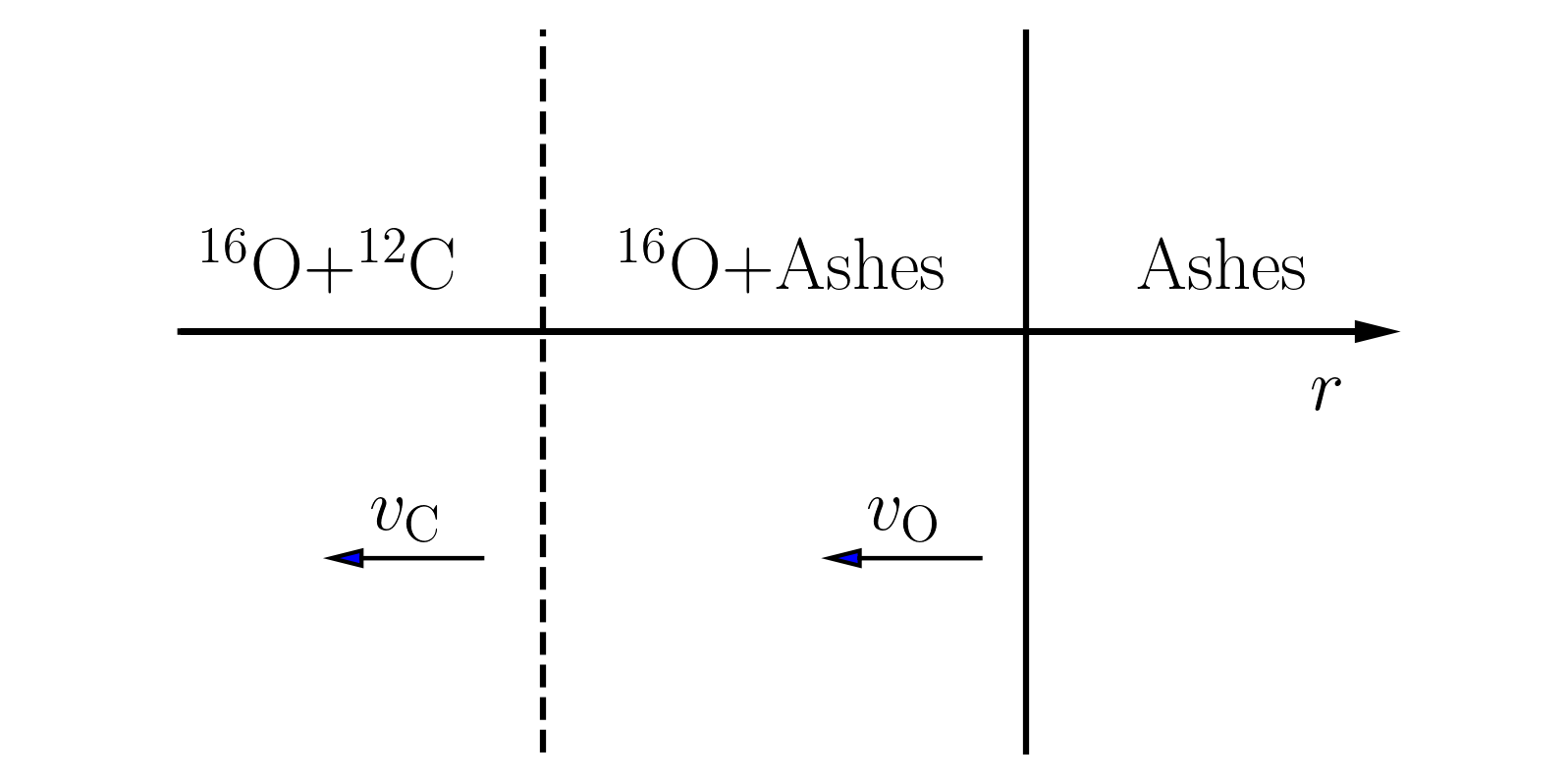}}
\caption{Schematic illustration of the physical scenario. (a) Schematic of an unburned island: the region enclosed by the arc and located above the main flame consists of unburned carbon and oxygen, while the region below contains the ashes produced by combustion. The dashed and solid curves represent the carbon and oxygen flames, respectively. The radial distribution of the isotopic composition in the blue box is shown in (b): the left side corresponds to unburned fuel, the right side consists of fully burned ash, and the middle region contains material where carbon has been consumed but oxygen remains unburned.}

\end{figure}


R--T instabilities in SNe~\uppercase\expandafter{\romannumeral1}a play a crucial role in shaping the morphology of the flame front as it propagates through the white dwarf. As buoyant plumes of hot ash rise and colder fuel sinks, these instabilities can fragment the flame surface and lead to the formation of discrete, isolated pockets of unburned fuel. \citet{shu_flame_2020} approximated such an isolated pocket, or ``fuel island,'' as a sphere surrounded by burned material, as illustrated in Figure~\ref{Fig.global_island}. The interior of the island consists of unburned carbon and oxygen, while the exterior is composed of ashes produced by previous carbon--oxygen burning.

At the relevant fuel densities, the ignition temperature of carbon is significantly lower than that of oxygen. As a result, carbon burning is triggered first, whereas oxygen burning experiences a finite induction delay. This naturally leads to a spatial separation between the carbon and oxygen flame fronts, with the oxygen flame trailing behind the carbon flame. This configuration is schematically illustrated in Figure~\ref{Fig.local_island}. 

Under the combined influence of flame stretching and turbulent mixing, the intermediate region composed of burned carbon ashes and unburned oxygen can be further elongated. Once the characteristic length of this oxygen-rich transition layer reaches a critical scale of order $\sim 10\ \rm{km}$, conditions become favorable for a Zel’dovich-gradient-type runaway. A spontaneous oxygen detonation can then be initiated within this region, which subsequently overtakes the carbon flame and triggers a secondary DDT in the carbon layer. This multi-stage process provides a plausible physical pathway for oxygen-flame-driven DDT in turbulent SNe~Ia environments.

\subsection{Equations and Nuclear Network}

In this work, we employ the hydrodynamics code Castro \citep{Almgren2020,2010ApJ...715.1221A,the_castro_development_team_2025_15785124} to perform direct numerical simulations of reactive flows. Castro is a fully compressible, Eulerian, adaptive mesh refinement (AMR) code designed for astrophysical applications, and solves the equations of compressible hydrodynamics coupled with nuclear reactions. The governing equations are the Euler equations with source terms for nuclear burning and thermal conduction, expressed as
\begin{align}
  &\frac{\partial \rho}{\partial t} + \nabla \cdot
    (\rho\bm{v}) = 0, \\
  &\frac{\partial (\rho X_{i})}{\partial t} + \nabla \cdot
    (\rho X_{i} \bm{v}) = \rho
    \dot{\omega}_{{i}}, \label{eq:mass} \\ 
  &\frac{\partial(\rho\bm{v})}{\partial t} + \nabla
    \cdot(\rho\bm{v}\bm{v}+p\bm{I}) = 0, 
    \label{eq:momentum} \\ 
  &\frac{\partial(\rho\epsilon)}{\partial t} + \nabla
    \cdot[\bm{v}(\rho\epsilon+p)] =  \rho \dot
    S\ + \nabla
    \cdot \left(k_{\text{th}}\nabla T\right). 
    \label{eq:energy} 
\end{align}
Here, $\rho$, $p$, $\bm{v}$, $\epsilon$, and $T$ denote the mass density, gas pressure, velocity, total specific energy, and temperature, respectively. $\bm{I}$ is the identity tensor. In Eq.~(\ref{eq:mass}), $X_{i}$ represents the mass fraction of the $i$th nuclear species, with $\sum_i X_i = 1$, and $\dot{\omega}_{i}$ is its corresponding production rate due to nuclear reactions. In Eq.~(\ref{eq:energy}), $\dot S$ denotes the specific nuclear energy generation rate, while $k_{\text{th}}$ is the thermal conductivity, evaluated following the formulation of \citet{Timmes_2000}. All thermodynamic quantities are computed using the Helmholtz equation of state (EOS) \citep{Timmes_2000_2}, which is well suited for degenerate, relativistic stellar matter.

\begin{figure}[htbp]
\centering
\includegraphics[width=0.47\textwidth]{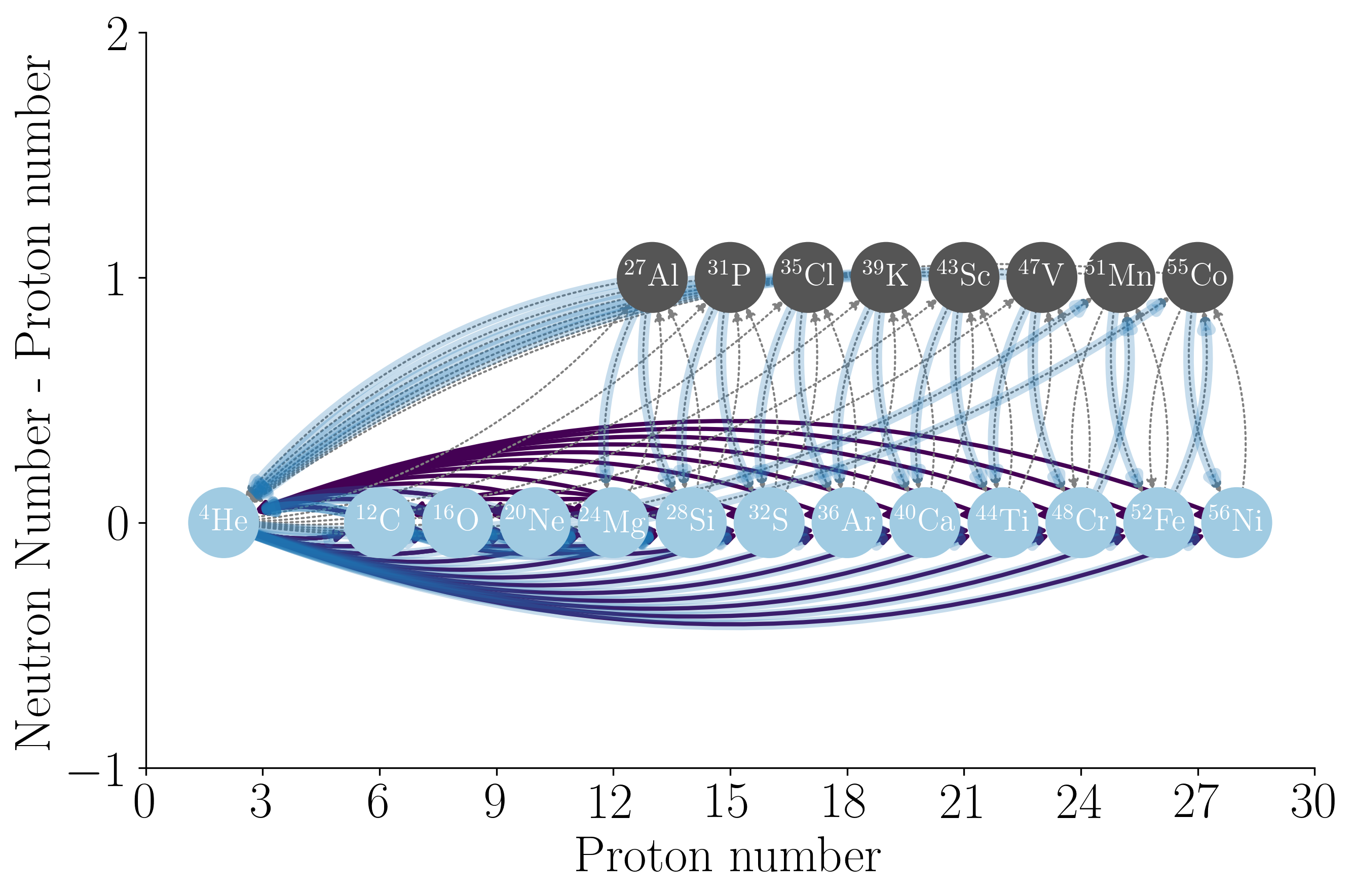}
\caption{The reaction network used in the simulation, includes the isotopes marked with blue circles; isotopes shown in gray circles are not explicitly included in the network.}
\label{Fig.network_ap13}
\end{figure}

The nuclear reaction network adopted in this study is the ``aprox13'' network\footnote{\url{https://cococubed.com/code_pages/burn_helium.shtml}}, as implemented through the pynucastro library \citep{pynucastro2,pynucastro,pynucastro_development_team_2025_15424953}. This network contains 13 isotopes: $^{4}$He, $^{12}$C, $^{16}$O, $^{20}$Ne, $^{24}$Mg, $^{28}$Si, $^{32}$S, $^{36}$Ar, $^{40}$Ca, $^{44}$Ti, $^{48}$Cr, $^{52}$Fe, and $^{56}$Ni (see Figure~\ref{Fig.network_ap13}). It includes the standard $\alpha$-chain reactions as well as the key heavy-ion fusion reactions ($^{12}$C+$^{12}$C, $^{12}$C+$^{16}$O, $^{16}$O+$^{16}$O). Approximate $(\alpha,\rm{p})(\rm{p},\gamma)$ reaction pathways are also incorporated, while intermediate isotopes such as $^{27}$Al, $^{31}$P, $^{35}$Cl, $^{39}$K, $^{43}$Sc, $^{47}$V, $^{51}$Mn, and $^{55}$Co are not explicitly tracked.

Screening corrections are applied using the prescription of \citet{wallace_thermonuclear_1982}, which combines the formulations of \citet{graboske_screening_1973}, \citet{jancovici_pair_1977}, \citet{alastuey_nuclear_1978}, and \citet{itoh_enhancement_1979}. Energy losses due to thermal neutrino processes---including pair, photo, plasma, and bremsstrahlung neutrino emission---are neglected in the present simulations for simplicity \citep{itoh_neutrino_1996}.

Nuclear reaction source terms are integrated using the VODE stiff ODE solver \citep{doi:10.1137/0910062}, and operator splitting between hydrodynamics and reactions is carried out via Strang splitting. This approach ensures both accuracy and stability in the presence of stiff nuclear burning timescales, while remaining computationally efficient for large-scale hydrodynamical simulations.


\section{1D results}\label{1D results}

\subsection{Simulation settings}
\label{sec.setting}

In the one-dimensional simulations, we model a planar configuration consisting of two adjacent layers: a $^{12}$C–$^{16}$O region (hereafter the CO region) and a $^{16}$O–ashes region (hereafter the OA region), as illustrated in Figure~\ref{Fig.local_island}. The setup follows the parameter choices of \citet{woosley_flames_2011} and \citet{shu_flame_2020}. The computational domain extends over $x \in [0,2a]$ with $a = 7.68\ \rm{km}$. The interface between the CO and OA regions is located at $x = a$.

The CO region represents unburned fuel and is initialized with a uniform density $\rho_0$ (to be specified in each simulation), temperature $T_0 = 6\times10^{8}\ \rm{K}$, and mass fractions $X_0(^{12}\rm{C}) = 50\%$ and $X_0(^{16}\rm{O}) = 50\%$. The corresponding pressure $p_0(\rho_0, T_0, \bm{X}_0)$ is determined from the EOS. The OA region represents material in which carbon has been consumed but oxygen has not yet fully burned. Its temperature varies linearly from $T_{1}=3.5\times10^{9}\ \rm{K}$ at the interface to $T_{2}=3.75\times10^{9}\ \rm{K}$ at the right boundary. The mass fractions are set to $X_1(^{16}\rm{O}) = 50\%$, $X_1(^{24}\rm{Mg}) = 25\%$, and $X_{1}(^{28}\rm{Si}) = 25\%$. For hydrostatic balance, the pressure in the OA region is set to $p = p_0$, and the density is then determined consistently from the EOS.

To avoid numerical artifacts at the interface, a smooth transition between the two regions is imposed using a hyperbolic tangent ($\tanh$) profile:
\begin{multline}
    T(x) = T_0 + \frac{1}{2}
        \left[1+\tanh\left(\frac{x-a}{w}\right) \right] \\
        \times \left[
        \frac{x-a}{a}(T_{2}-T_{1}) + T_{1}-T_{0}
        \right] ,
\end{multline}

\begin{equation}
    \bm{X}(x) = \bm{X}_{0}+\frac{1}{2}
    \left[1+\tanh\left(\frac{x-a}{w}\right)\right](\bm{X}_1-\bm{X}_0),
\end{equation}
where $w$ characterizes the initial thickness of the carbon deflagration front and its value is $30.72\ \rm{m}$ unless otherwise noted (c.f. \S\ref{sec:width}).

The computational grid consists of 4096 uniform base cells, with Adaptive Mesh Refinement (AMR) enabled up to level~6. This corresponds to a finest spatial resolution of $\Delta x_{\rm min} = 5.86\ \rm{cm}$. Outflow and inflow boundary conditions are applied at the left and right boundaries, respectively. Each simulation is evolved for a total physical time of $1.8\ \rm{ms}$, with a Courant--Friedrichs--Lewy (CFL) number of 0.3 to ensure numerical stability.

\subsection{Fiducial 1D Simulation}
\begin{figure*}[htbp]
\centering
\includegraphics[width=0.83\textwidth]{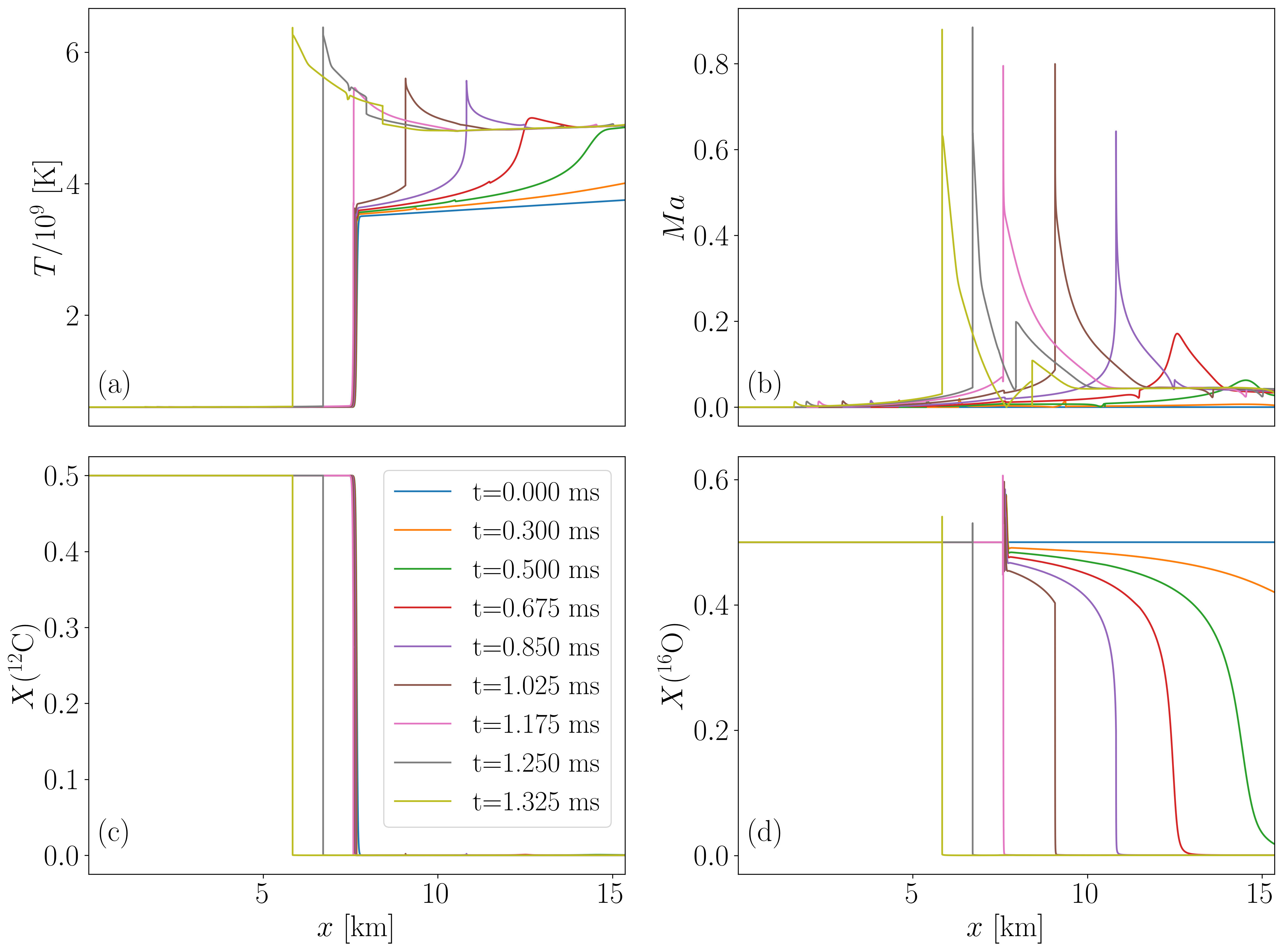}
\caption{1D simulation at $\rho_{0}=3.5\times 10^{7}\ \rm{g\ cm^{-3}}$ showing successful DDT: the evolution of (a) temperature, (b) Mach number, (c) $^{12}\rm{C}$ mass fraction, and (d) $^{16}\rm{O}$ mass fraction, respectively.}
\label{Fig.rho=3.5}
\end{figure*}

\begin{figure}[htbp]
\centering
\includegraphics[width=0.45\textwidth]{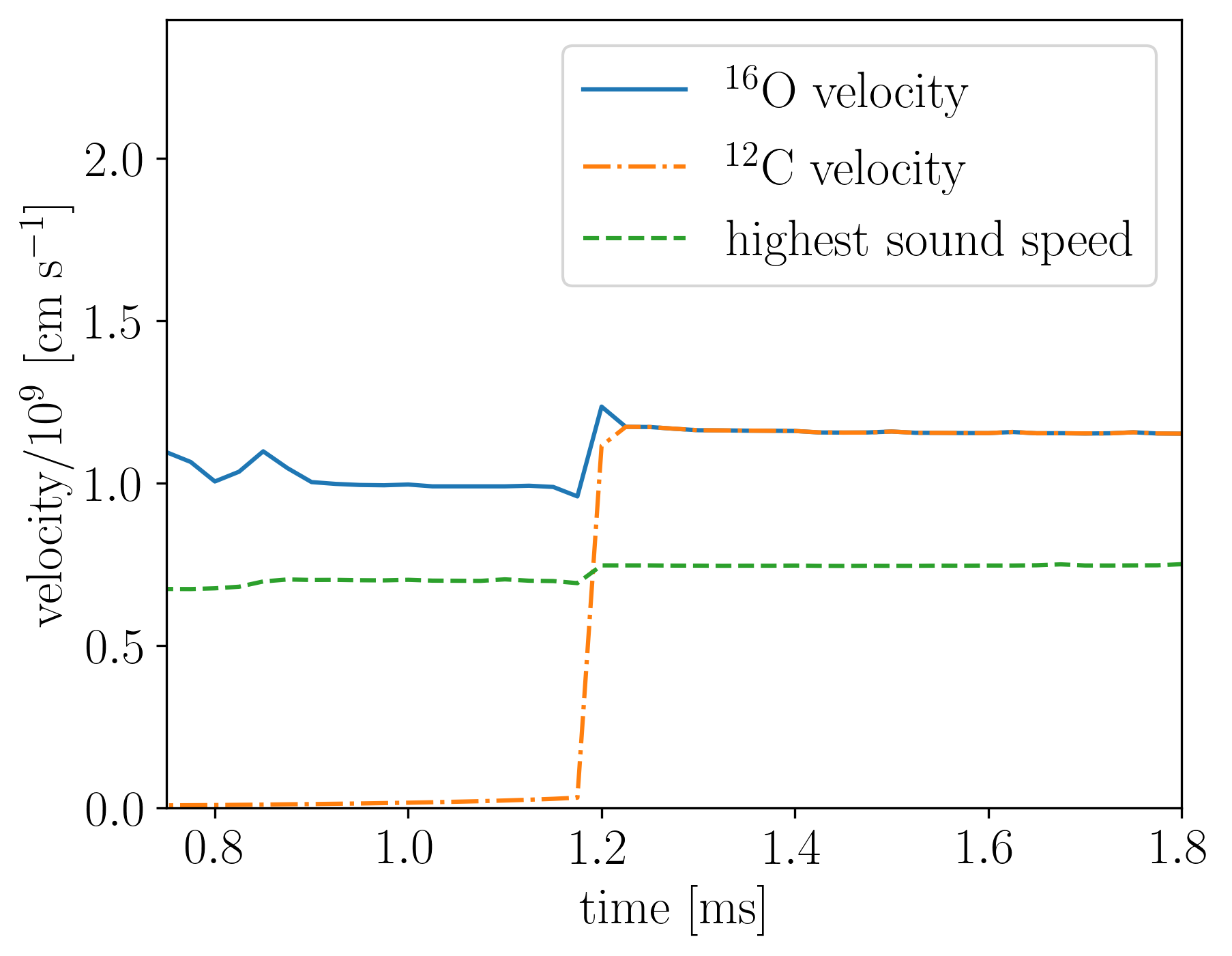}
\caption{Comparison of the carbon and oxygen flame velocities with the maximum sound speed. Throughout the detonation, the flame velocities consistently exceeds the highest sound speed in the domain.}
\label{Fig.vel}
\end{figure}

\begin{figure*}[htbp]
\centering
\includegraphics[width=0.83\textwidth]{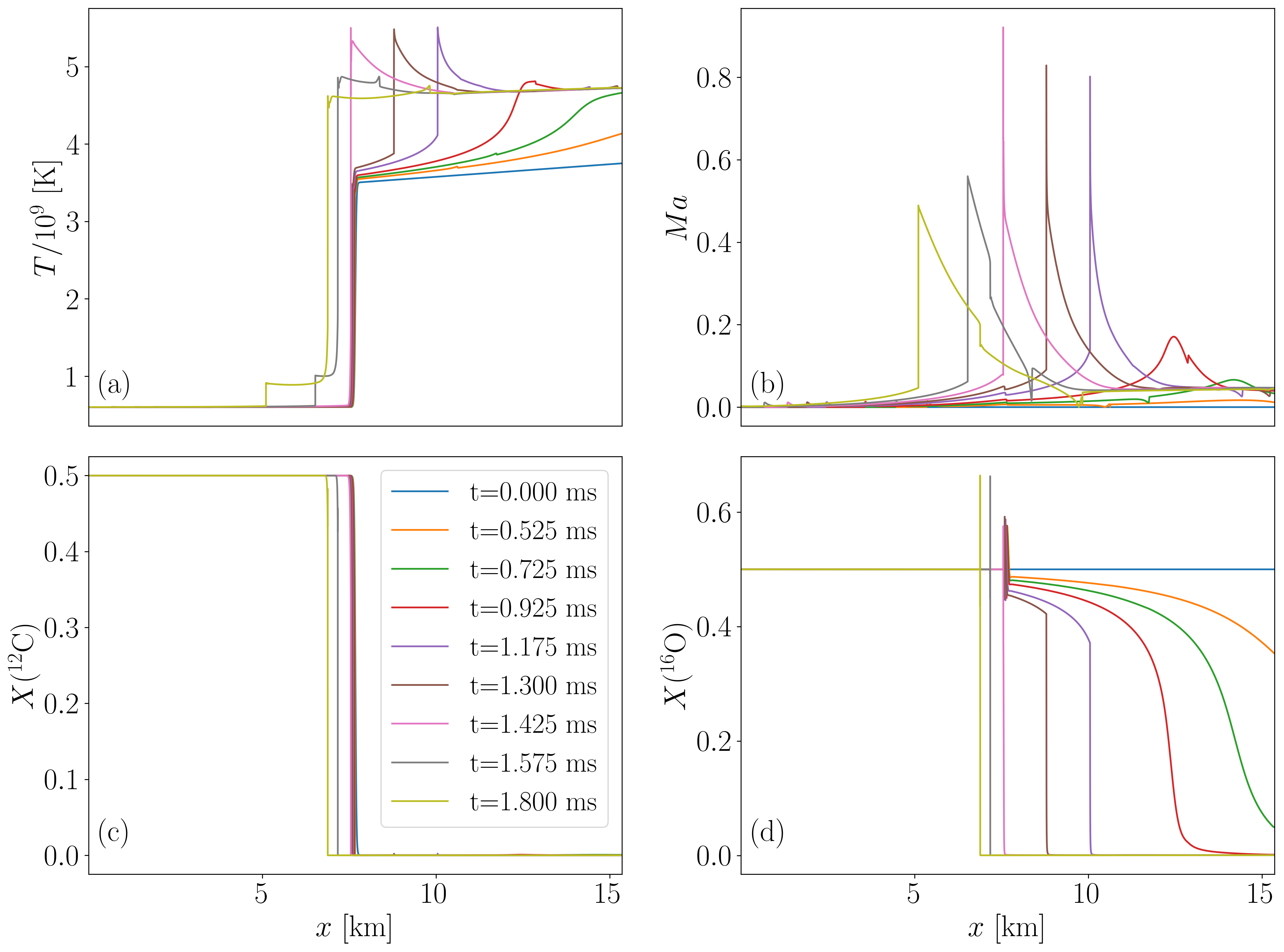}
\caption{1D simulation at $\rho_{0}=3\times 10^{7}\ \rm{g\ cm^{-3}}$. The upper left, upper right, lower left, and lower right panels show the evolution of temperature, Mach number, $^{12}\rm{C}$ mass fraction, and $^{16}\rm{O}$ mass fraction, respectively.}
\label{Fig.rho=3.0}
\end{figure*}

Prior to running the simulations, we estimated the maximum densities attainable under the adopted initial conditions (Appendix~\ref{density estimate}). This estimate shows that the density must not exceed $3.6\times10^{7}\ \rm{g\ cm^{-3}}$ in order to maintain the desired flame separation. Based on this result, we perform a fiducial one-dimensional simulation of DDT at $\rho_0 = 3.5\times10^{7}\ \rm{g\ cm^{-3}}$, as shown in Figure~\ref{Fig.rho=3.5}. The figure illustrates the temporal evolution of the temperature, Mach number, and the mass fractions of $^{12}$C and $^{16}$O.

At $t = 0\ \rm{ms}$, the velocity is zero throughout the domain, and a temperature gradient of approximately $0.25\times10^{9}\ \rm{K}$ is imposed on the right-hand side. Burning is first ignited in this region (see panels at $t = 0.3\ \rm{ms}$ and $t = 0.5\ \rm{ms}$), which subsequently drives the formation of a shock wave. As the evolution proceeds, the shock and the flame front mutually reinforce each other, leading to a progressive increase in the flame-front temperature. By $t = 1.025\ \rm{ms}$, a sharp temperature and pressure discontinuity emerges, marking the onset of DDT. At this point, the flame-front temperature reaches approximately $5.5\times10^{9}\ \rm{K}$.

Although the post-shock Mach number remains subsonic, the flame front itself propagates at a speed of $10^{10}\ \rm{cm\ s^{-1}}$ (Figure~\ref{Fig.vel}), far exceeding the maximum sound speed in the domain. The oxygen flame remains tightly coupled to the leading shock and propagates steadily. Around $t = 1.175\ \rm{ms}$, the oxygen detonation front overtakes the carbon deflagration front, triggering a DDT in the carbon layer and producing a stronger detonation wave. With additional unburned carbon available, this secondary detonation reaches a higher peak temperature and Mach number. The resulting detonation then propagates stably at approximately $1.15\times10^{10}\ \rm{cm\ s^{-1}}$, consistent with the characteristic speed of a carbon–oxygen detonation \citep{khokhlov_deflagrationdetonation_1997}. Finally, a resolution study confirms that an AMR level of~6 is sufficient to ensure numerical convergence (Appendix~\ref{Convergence Test}).

\subsection{Minimum Density for 1D DDT}

To investigate the density conditions under which carbon detonation can occur, we perform a series of one-dimensional simulations varying the initial fuel density $\rho_{0}$ toward lower values. The results indicate that successful carbon DDT occurs only for $\rho_{0} \gtrsim 3.1\times10^{7}\ \rm{g\ cm^{-3}}$. Figure~\ref{Fig.rho=3.0} shows the representative case with $\rho_{0}=3\times10^{7}\ \rm{g\ cm^{-3}}$. During the early evolution ($t \lesssim 1.025\ \rm{ms}$), the oxygen flame behaves similarly to the fiducial run: it gradually generates a shock wave through the Zel’dovich gradient mechanism, which subsequently couples to the flame and accelerates, forming an oxygen detonation. However, this oxygen detonation fails to trigger a secondary carbon detonation. When the oxygen detonation front reaches the carbon deflagration front, the shock and flame fail to couple; the shock continues to propagate at high speed, while both carbon and oxygen continue to burn in a subsonic deflagration mode.

\begin{figure}[htbp]
\centering
\includegraphics[width=0.45\textwidth]{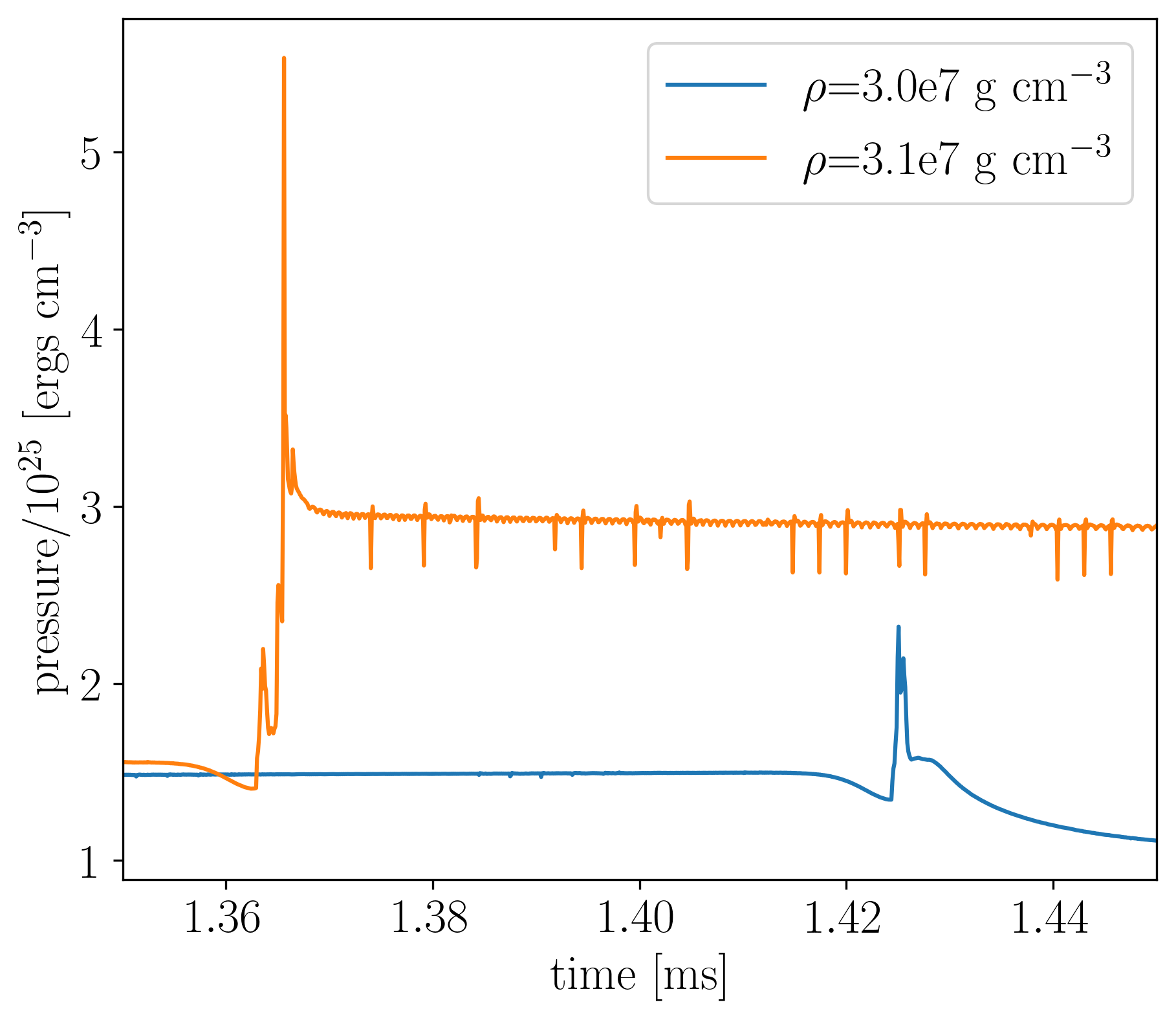}
\caption{Evolution of the shock pressure as the oxygen detonation front catches up with the carbon deflagration front at $\rho_{0}=3.1\times 10^{7}\ \rm{g\ cm^{-3}}$ and $3\times 10^{7}\ \rm{g\ cm^{-3}}$.}
\label{Fig.3.0_3.1}
\end{figure}

To understand the failure of DDT at lower densities, we examine the temporal evolution of the shock pressure for $\rho_{0}=3.0$ and $3.1\times10^{7}\ \rm{g\ cm^{-3}}$ (Figure~\ref{Fig.3.0_3.1}). At $\rho_{0}=3.1\times10^{7}\ \rm{g\ cm^{-3}}$, the pressure exhibits sustained oscillatory growth, reaching values up to $5\times10^{25}\ \rm{erg\ cm^{-3}}$ before relaxing to $\sim3\times10^{25}\ \rm{erg\ cm^{-3}}$ following carbon detonation. In contrast, at $\rho_{0}=3.0\times10^{7}\ \rm{g\ cm^{-3}}$, the pressure peaks at $\sim2.3\times10^{25}\ \rm{erg\ cm^{-3}}$ and then rapidly declines below $2\times10^{25}\ \rm{erg\ cm^{-3}}$. The reduced nuclear reaction rates at lower densities delay carbon ignition, preventing the flame from coupling with the shock before decoupling occurs. As a result, the system fails to transition to detonation and reverts to a deflagration regime.

\begin{figure}[htbp]
\centering
\includegraphics[width=0.45\textwidth]{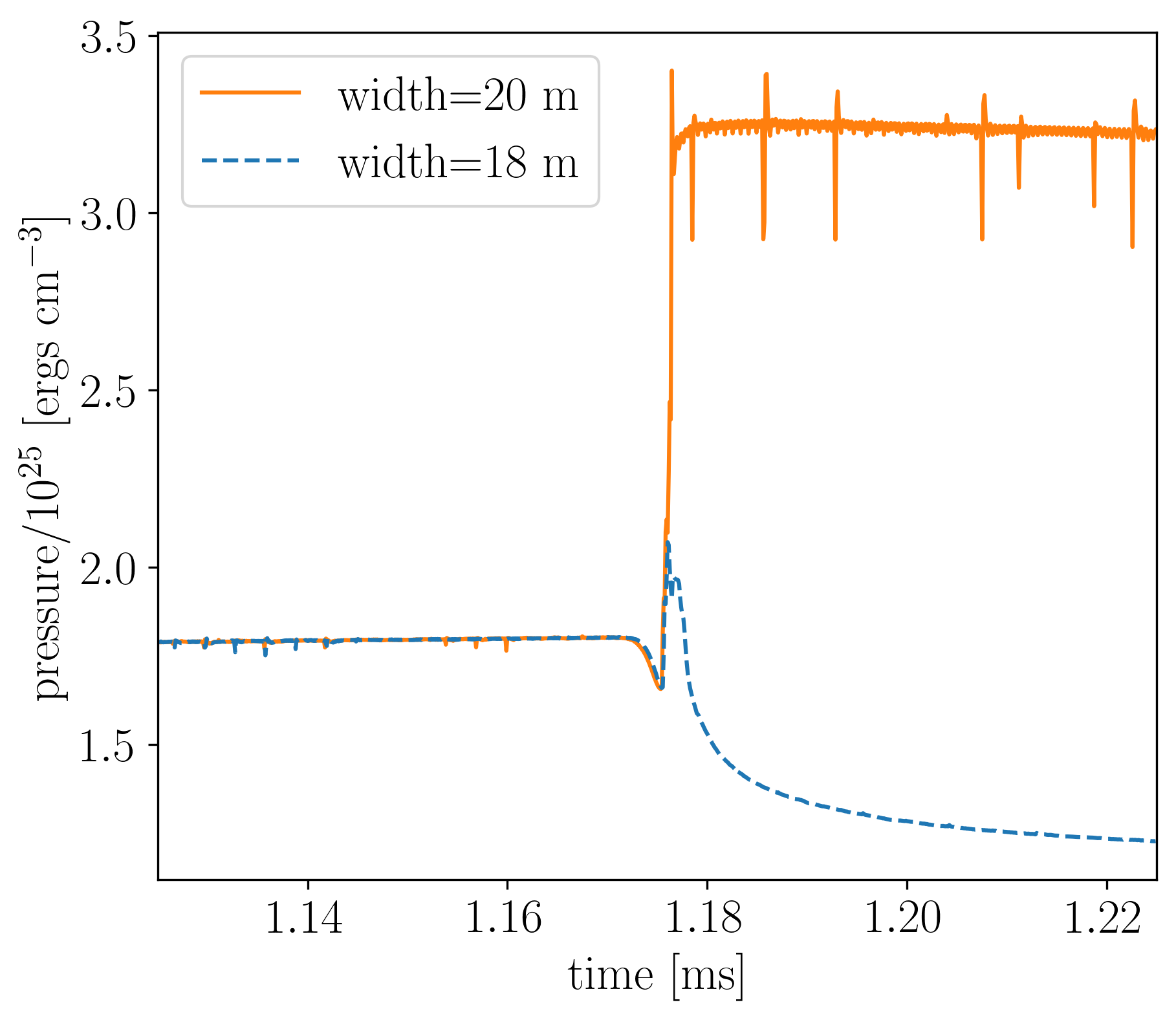}
\caption{Evolution of the shock pressure as the oxygen detonation front catches up with the carbon deflagration front at thickness equal to 18 and 20 m.}
\label{Fig.width}
\end{figure}
\subsection{Critical Thickness of the Carbon Flame Front for 1D DDT}\label{sec:width}

\begin{figure*}[htbp]
\centering
\includegraphics[width=0.83\textwidth]{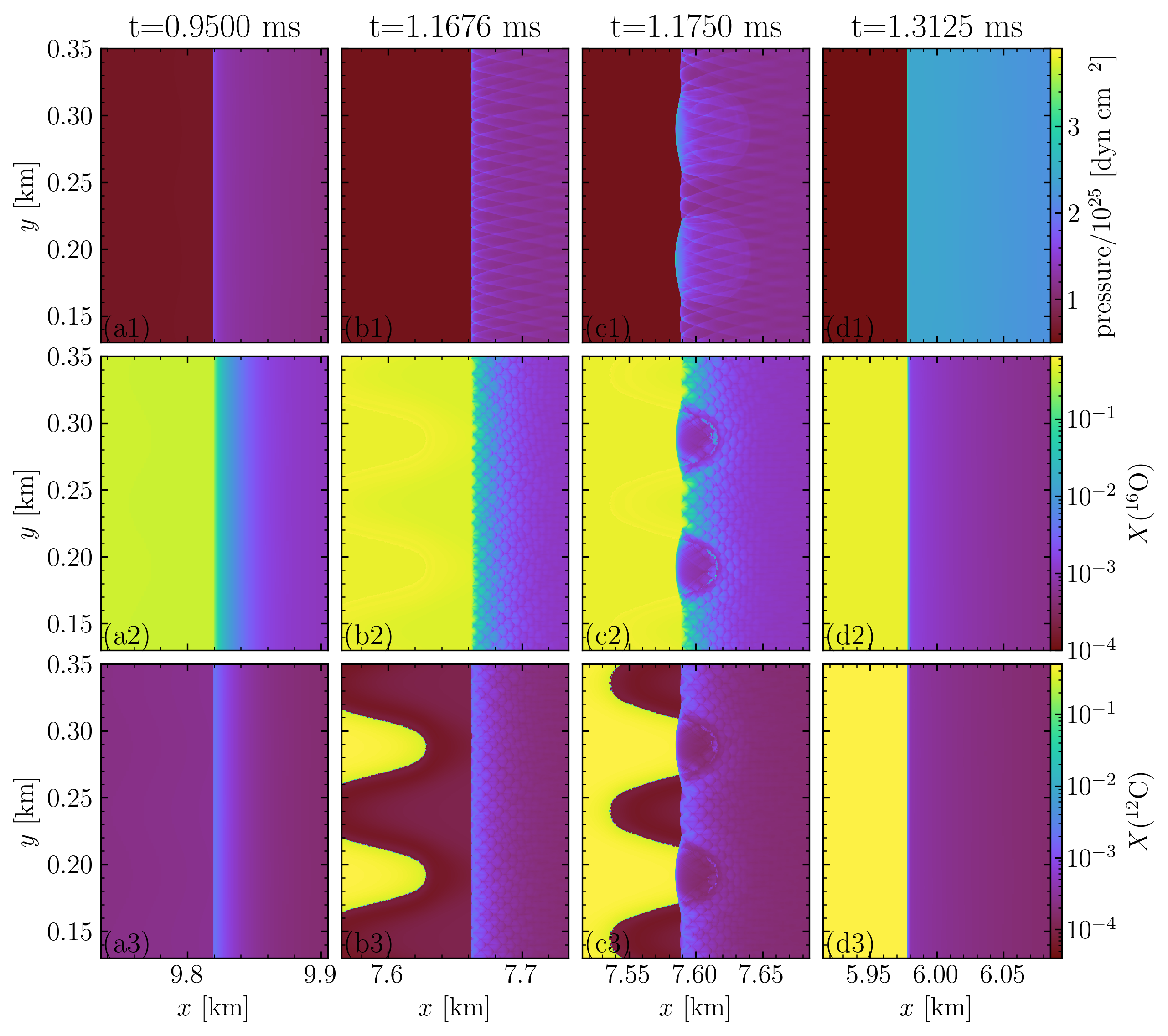}
\caption{2D simulations at density of $3.5\times10^{7}\ \rm{g\ cm^{-3}}$. From top to bottom, the rows show the pressure, $^{16}$O mass fraction, and $^{12}$C mass fraction near the flame front, respectively. From left to right, the columns show snapshots at $t=0.95$ ms, 1.1676 ms, 1.175 ms, and 1.3125 ms, respectively.}
\label{Fig.2d}
\end{figure*}

\begin{figure*}[htbp]
\centering
\includegraphics[width=0.83\textwidth]{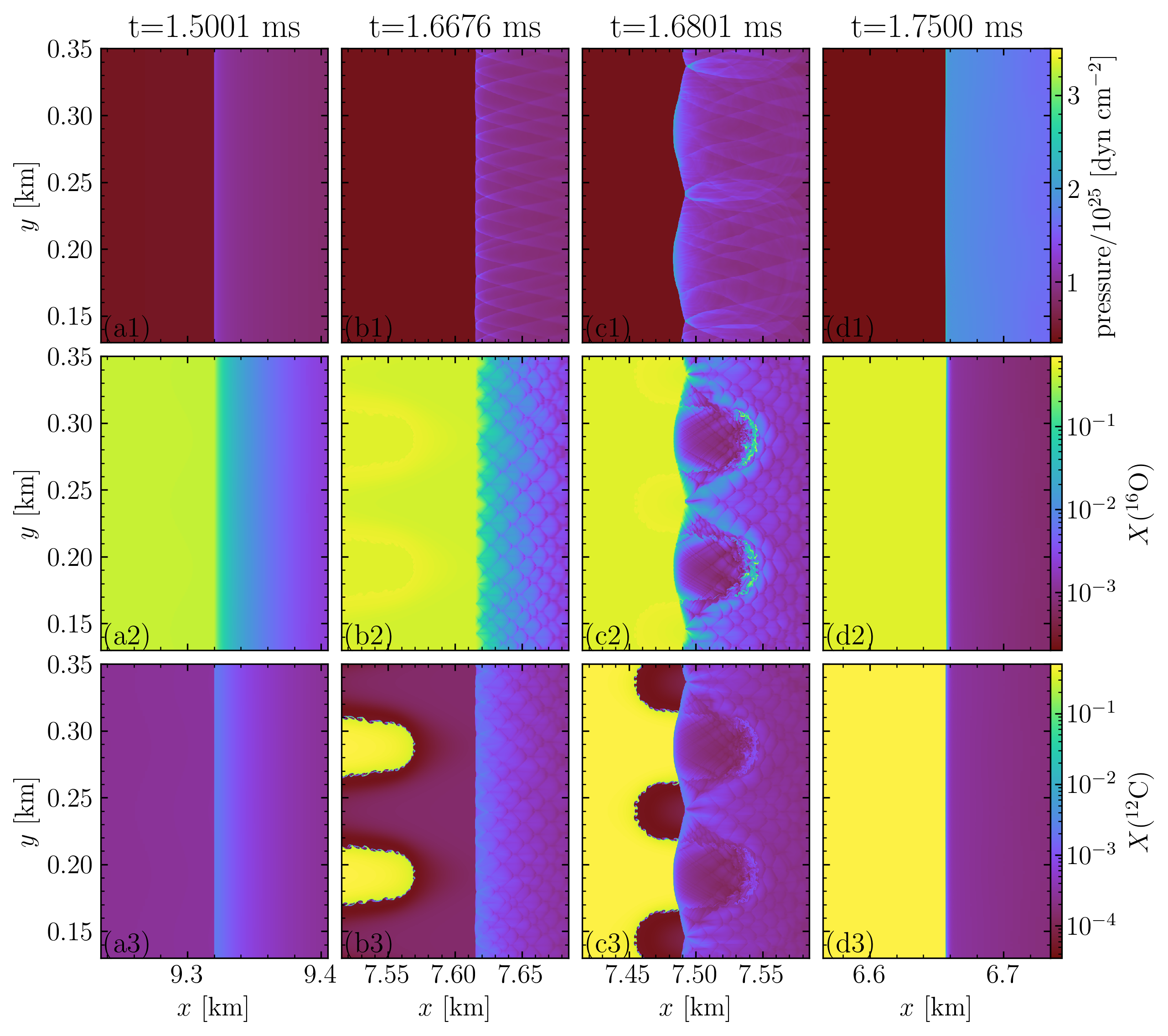}
\caption{2D simulations at density of $2.7\times10^{7}\ \rm{g\ cm^{-3}}$. 
 From top to bottom, the rows show the pressure, $^{16}$O mass fraction, and $^{12}$C mass fraction near the flame front, respectively. From left to right, the columns show snapshots at $t= 1.5$ ms, 1.6676 ms, 1.68 ms, and 1.755 ms, respectively.}
\label{Fig.2d_2.7e7}
\end{figure*}

In the one-dimensional simulations, we also find that the thickness of the carbon deflagration front plays a crucial role in determining whether DDT occurs. Here, the flame thickness is defined as the distance between the positions where the $^{12}$C mass fraction decreases from 49\% to 1\%. If the flame is too thin, the shock and carbon flame do not have sufficient time to couple, and the transition fails.

To quantify this effect, we vary the parameter $w$ introduced in \S\ref{sec.setting}, which controls the initial flame thickness, while fixing the density at $\rho_{0}=3.5\times10^{7}\ \rm{g\ cm^{-3}}$. The results indicate that carbon DDT can only be triggered when the flame thickness exceeds $\sim$20~m (Figure~\ref{Fig.width}). This threshold is much larger than the laminar carbon flame thickness, which is only of order 1~cm under quiescent conditions \citep{niemeyer_small-scale_1999, woosley_flames_2011}.


In a turbulent environment the Reynolds number can reach approximately \(10^{13}\gg1\) \citep{shu_flame_2020}, so the flame structure may be significantly broadened. Following the discussion in Appendix~\ref{density estimate}, the turbulent carbon flame width can be estimated as \(\delta_{\rm C}\sim u'\,\tau_{\rm C}=u'\frac{\delta_{\rm C,l}}{u_{\rm C,l}}\), where \(\delta_{\rm C}\) is the turbulent flame width, \(u'\) is the turbulent velocity, \(\tau_{\rm C}\) is the carbon burning timescale, and \(\delta_{\rm C,l}\) and \(u_{\rm C,l}\) are the laminar flame thickness and laminar flame speed respectively. The laminar carbon flame speed is \(u_{\rm C,l}\sim10^{4}\ \mathrm{cm\ s^{-1}}\) \citep{niemeyer_small-scale_1999}. If the turbulent velocity attains \(u'\sim1000\ \mathrm{km\ s^{-1}}\), then \(\delta_{\rm C}\sim 10^{8}\frac{\delta_{\rm C,l}}{10^{4}}\sim 10^{4}\,\delta_{\rm C,l}\). Thus, for typical laminar thicknesses in the cm range, the turbulence-broadened flame thickness can reach the order of \(10^{2}\ \mathrm{m}\). Therefore, the flame thicknesses required for carbon DDT in our simulations (\(\sim 100\ \mathrm{m}\)) are physically plausible in realistically turbulent white-dwarf interiors. In addition, because the turbulence-broadened widths satisfy \(\delta_{\rm C}\gg\Delta x_{\min}\) (where \(\Delta x_{\min}\) denotes the minimum grid spacing in our AMR setup), numerical dissipation associated with the finite-resolution scheme is small compared with the physical broadening; hence the assumption of a turbulence-broadened flame is consistent with our numerical resolution.

\section{2D results}\label{2D results}
\subsection{simulation settings}


Building on the one-dimensional configuration described in \S\ref{sec.setting}, we extend the setup to two dimensions. The computational domain is identical in the $x$ direction, while in the $y$ direction it spans $0.48\ \rm{km}$. The maximum AMR level remains 6, and periodic boundary conditions are imposed along the $y$ direction. 

To introduce a perturbation at the interface between the CO and OA regions, the interface position $x_{\rm mid}$ is modulated as a function of $y$:
\begin{equation}
    x_{\rm mid}(y) = a + \delta \cos\left(\frac{2\pi y}{\lambda}\right),
\end{equation}
where $\delta = 32\ \rm{m}$ and $\lambda = 96\ \rm{m}$. This sinusoidal perturbation breaks the initial planar symmetry and seeds multidimensional effects. All other simulation parameters are identical to those adopted in the one-dimensional calculations.

\subsection{Fiducial 2D simulation}


In two-dimensional runs we also observe a deflagration–to–detonation transition (DDT) at
\(\rho_{0}=3.5\times10^{7}\ \mathrm{g\ cm^{-3}}\).  Figure~\ref{Fig.2d} shows the formation
sequence of the 2D carbon detonation.  At \(t=0.95\ \mathrm{ms}\) an oxygen detonation front
is already present, but it remains essentially quasi–one–dimensional and has not yet developed
a clear lateral cellular structure.  After further evolution, by \(t=1.1676\ \mathrm{ms}\),
as the oxygen front approaches the carbon deflagration, a distinct two-dimensional detonation
morphology appears: regular transverse waves and triple-point structures form along the front,
and the mass fractions of \(^{16}\mathrm{O}\) and \(^{12}\mathrm{C}\) exhibit clear cell-like
patterns in the wake.

At \(t=1.175\ \mathrm{ms}\) the oxygen detonation overtakes the carbon deflagration and triggers
DDT in the carbon layer.  This interaction disrupts the regular transverse pattern, producing a
highly irregular front and leaving small pockets of unburned fuel, which indicates a transiently
unstable detonation state.  Despite this transient instability, a self-sustained carbon detonation
is initiated.  As the coupled carbon–oxygen detonation propagates further, by \(t=1.3125\ \mathrm{ms}\)
the front has strengthened and smoothed into a nearly planar configuration closely resembling a
Chapman–Jouguet (CJ) detonation, demonstrating the ability of the C–O detonation to regain stability
after the transition.

\begin{figure}[htbp]
\centering
\includegraphics[width=0.45\textwidth]{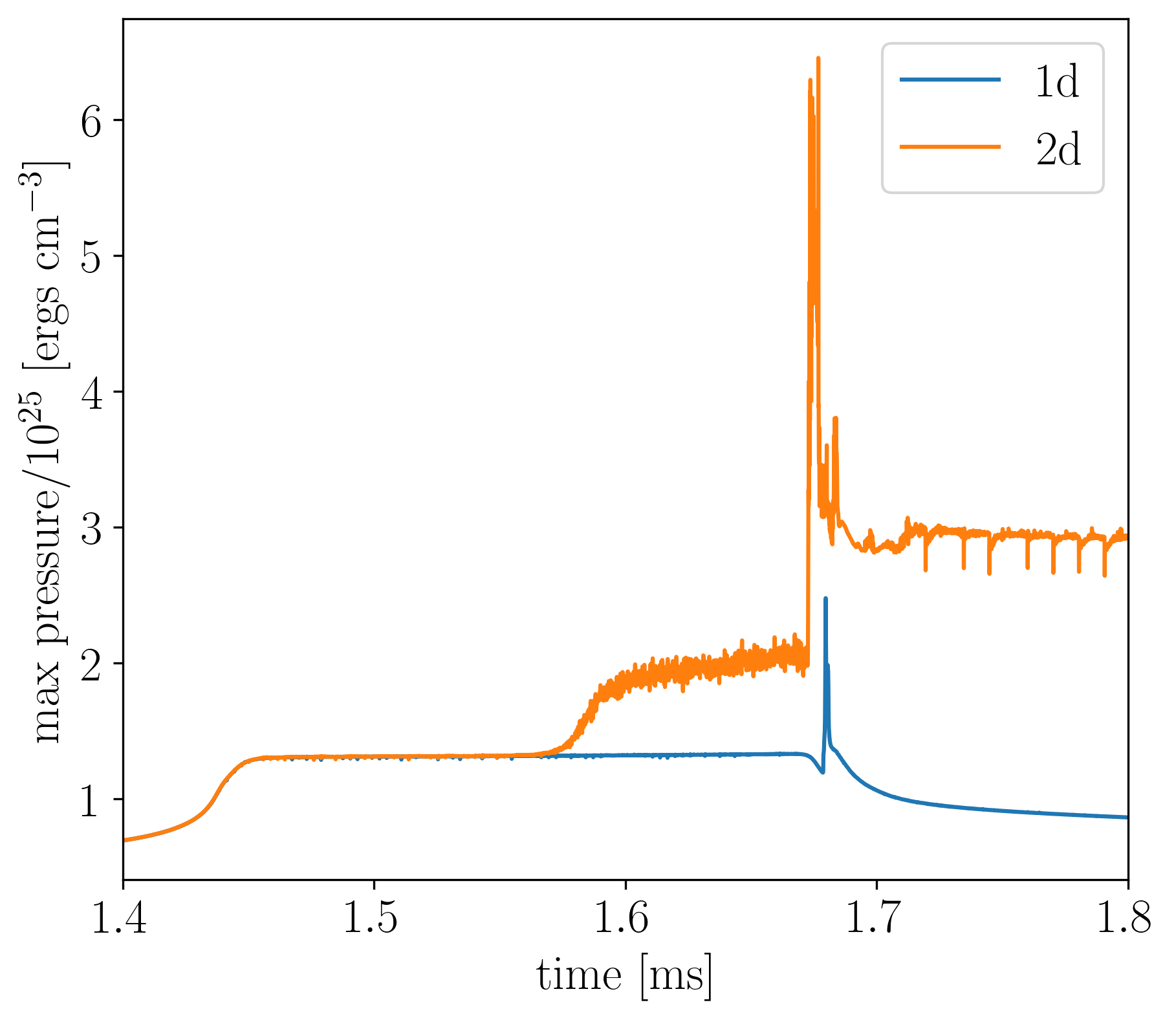}
\caption{Evolution of the max pressure of 1D and 2D simulations at $2.7\times 10^{7}\ \rm{g\ cm^{-3}}$.}
\label{Fig.2.7_1d_2d}
\end{figure}

\subsection{DDT at lower density}

We also performed a two-dimensional simulation at an upstream density of
\(\rho_{0}=2.7\times10^{7}\ \mathrm{g\ cm^{-3}}\).  The results, shown in
Figure~\ref{Fig.2d_2.7e7}, display a qualitative difference from the 1D case:
a self-sustained carbon detonation appears in 2D although it is absent in 1D.
This demonstrates the important role of multidimensional structure in enabling
DDT under conditions where DDT fails in 1D simulations.

The oxygen detonation evolves from a quasi–1D front into a clearly
two–dimensional cellular front between \(t=1.50\ \mathrm{ms}\) and
\(t=1.6676\ \mathrm{ms}\).  At \(t=1.6676\ \mathrm{ms}\) the characteristic
cell size in the \(X(^{16}\mathrm{O})\) field is noticeably larger than that
seen in Figure~\ref{Fig.2d}, consistent with slower reaction rates at the
lower density and with an enhanced propensity for transverse instability.
By \(t=1.68\ \mathrm{ms}\) the oxygen front has overtaken the carbon flame,
leaving a larger pocket of unburned fuel than in the \(\rho_{0}=3.5\times10^{7}\)
case.  Despite this increased irregularity, the interaction between oxygen and
carbon burning ultimately leads (by \(t=1.75\ \mathrm{ms}\)) to a well-coupled,
stable carbon–oxygen detonation in 2D.  We note, however, that the lower-bound
for DDT remains constrained by the minimum density at which oxygen burning can
sustain a detonation \citep{woosley_flames_2011}.

To quantify why carbon detonation is achieved in 2D but not in 1D, we tracked
the maximum pressure \(P_{\max}(t)\) in each simulation.  Figure~\ref{Fig.2.7_1d_2d}
compares \(P_{\max}(t)\) for the 1D and 2D runs.  When the oxygen detonation
first forms the two runs have nearly identical peak pressures (quasi–1D stage).
Beginning at \(t\approx1.57\ \mathrm{ms}\), however, the development of
two–dimensional structures produces localized pressure spikes that exceed the
1D values.  These high-pressure spots are the sites where induction times in
the carbon-rich mixture are strongly reduced, and they play a decisive role in
seeding localized carbon ignition and the subsequent DDT.

The localized pressure amplification arises naturally from triple-point
collisions and the constructive interference of transverse waves (see
Figure~\ref{Fig.2d_2.7e7}).  Such mechanisms are absent in strictly 1D
geometries, which explains the divergent outcomes between 1D and 2D at the
same upstream density.  In short, the 2D run shows that multidimensional shock
interactions can produce transient overpressure regions sufficient to trigger
carbon detonation even when global, one-dimensional conditions are marginal. This behaviour is consistent with previous findings in chemistry burning \citep{gamezo_flame_2008,oran_origins_2007}.

\section{Summary and the Future works}
\label{summary}

In this paper we performed direct numerical simulations of oxygen-flame-driven
deflagration–to–detonation transition (DDT) in Type Ia supernova conditions using
the Castro hydrodynamics code coupled with the ``aprox13'' nuclear reaction
network.
Our main findings are summarized as follows. We find that oxygen-flame-driven DDT can
occur via the Zel'dovich gradient mechanism when a sufficiently large separation
(roughly $\sim 10\ \mathrm{km}$) exists between the carbon and oxygen burning layers,
so that an oxygen detonation front can overtake the carbon flame and induce a localized
carbon runaway. From systematic 1D simulations we identify a window of upstream fuel
densities and a minimum carbon-flame thickness required for carbon DDT: carbon DDT
occurs only for $\rho_{0}\simeq(3.1$--$3.6)\times10^{7}\ \mathrm{g\ cm^{-3}}$ and for
carbon flame thicknesses $\gtrsim 20\ \mathrm{m}$. Two-dimensional calculations show
that the multidimensional structure of the oxygen detonation (transverse waves,
triple points and associated shock focusing) can enhance the formation of carbon
detonation at lower densities than predicted by strictly one-dimensional models,
highlighting the importance of higher-dimensional simulations for capturing the
relevant shock–induction coupling.
These results support the physical possibility of oxygen-flame-driven DDT in SNe~Ia
under suitable local conditions, while underscoring the sensitivity of the process
to upstream density and the spatial separation between fuel layers. In particular,
multidimensional shock interactions can locally amplify pressure and reduce induction
times, thereby relaxing one-dimensional thresholds for DDT.

For future work we plan two complementary efforts. First, we will carry out
higher-fidelity multi-dimensional fluid simulations to assess whether Rayleigh–Taylor
and related hydrodynamic instabilities can generate sufficiently strong vortical
motions to produce the required spatial separation (of order $\sim 10\ \mathrm{km}$)
between carbon and oxygen flames. Second, to make large-domain, high-resolution
multi-dimensional studies computationally tractable, we will develop and apply
data-driven acceleration techniques (e.g., neural-network-based surrogates and
emulators) following the approaches of \citet{fan_neural_2022} and
\citet{zhang_deep_2025}, with the ultimate goal of testing the robustness of
oxygen-flame-driven DDT in realistic, high-dimensional SN~Ia environments.
\\

Computational Expenses: the 1D simulations are performed on Intel Xeon ICX Platinum 8358 CPUs, cost approximately 150 core hours per run. The 2D simulations are performed on NVIDIA RTX 4090 GPUs, with a computational cost of approximately 2,000 GPU hours per run.

\software{Castro \citep{Almgren2020,2010ApJ...715.1221A,the_castro_development_team_2025_15785124} , pynucastro \citep{pynucastro2,pynucastro,pynucastro_development_team_2025_15424953}, Microphysics \citep{amrex_astro_microphysics_development_tea_2025_15782721}, AMReX \citep{AMReX_JOSS}, yt \citep{turk_yt_2011}.}

\begin{acknowledgments}
\noindent This work is supported by the National Natural Science Foundation of China under Grants 12305246 and the Fundamental Research Funds for the Central Universities.
\end{acknowledgments}

\appendix
\section{Upper-Bound Estimate for the Density}
\label{density estimate}


Based on the initial setup in \ref{sec.setting}, the $^{16}\mathrm{O}$–ashes region has a length of approximately $7.5\ \mathrm{km}$. According to \cite{ropke_flamedriven_2007}, turbulent velocities inside the white dwarf can reach up to $1000\ \mathrm{km\ s^{-1}}$. Combining this with the results of \cite{shu_flame_2020}, within vortices with radii larger than $100\ \mathrm{km}$ the carbon flame velocity can be accelerated by up to a factor of $1.5$, i.e. reaching $\sim 1500\ \mathrm{km\ s^{-1}}$.

Following the order-of-magnitude estimate of \cite{woosley_flames_2011}, we take the turbulent flame width to be
$\delta_{\rm C\text{-}O}\sim u'\,\tau_{\rm C\text{-}O}$,
assuming that turbulent eddies can stretch the flame front over a typical distance during a nuclear burning timescale. In our context, $\delta_{\rm C\text{-}O}$ is the turbulence-broadened separation between the carbon and oxygen flames, $u'$ is the turbulent velocity, and $\tau_{\rm C\text{-}O}$ is the induction (delay) time associated with delayed carbon–oxygen burning. This implies that a delay time of at least $5\ \mathrm{ms}$ is required to establish the initial separation used in our setup.

\begin{figure}[htbp]
\centering
\includegraphics[width=0.5\textwidth]{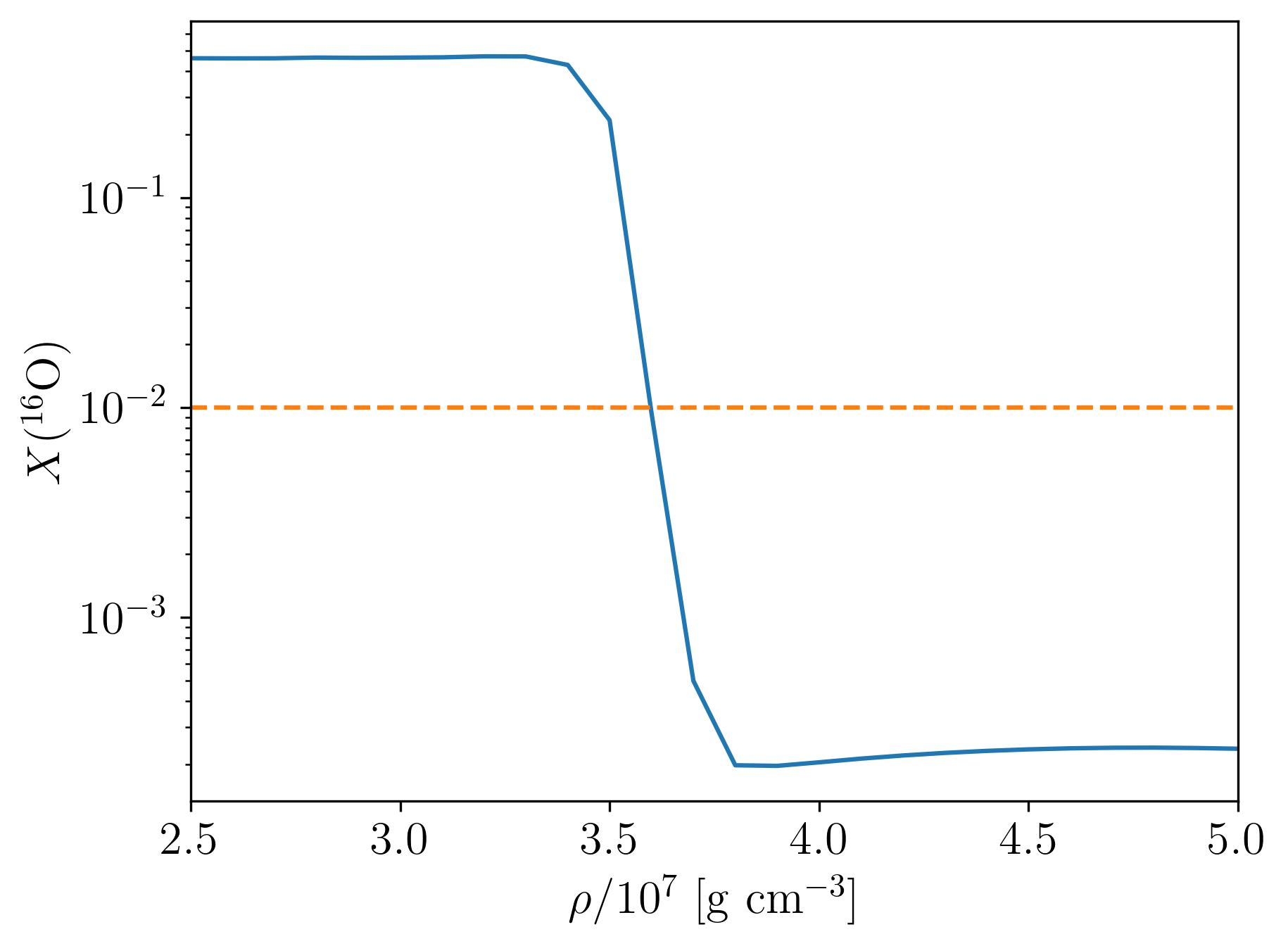}
\caption{Residual $X(^{16}\rm{O})$ at $t=5\ \rm{ms}$ in 1D deflagration simulations at different densities.}
\label{Fig.delay_ignition}
\end{figure}

Therefore, in the 1D deflagration simulations we vary the density to examine oxygen burning on a $5\ \mathrm{ms}$ timescale. If the oxygen mass fraction remains above $0.01$ after $5\ \mathrm{ms}$, we regard this as indicating that the delay time at that density exceeds $5\ \mathrm{ms}$. As shown in Figure~\ref{Fig.delay_ignition}, even under these idealized conditions the highest density that meets the $5\ \mathrm{ms}$ requirement is no more than $3.6\times10^{7}\ \mathrm{g\ cm^{-3}}$.

\FloatBarrier

\section{1D Convergence Test}
\label{Convergence Test}

\begin{figure}[htbp]
\centering
\includegraphics[width=0.45\textwidth]{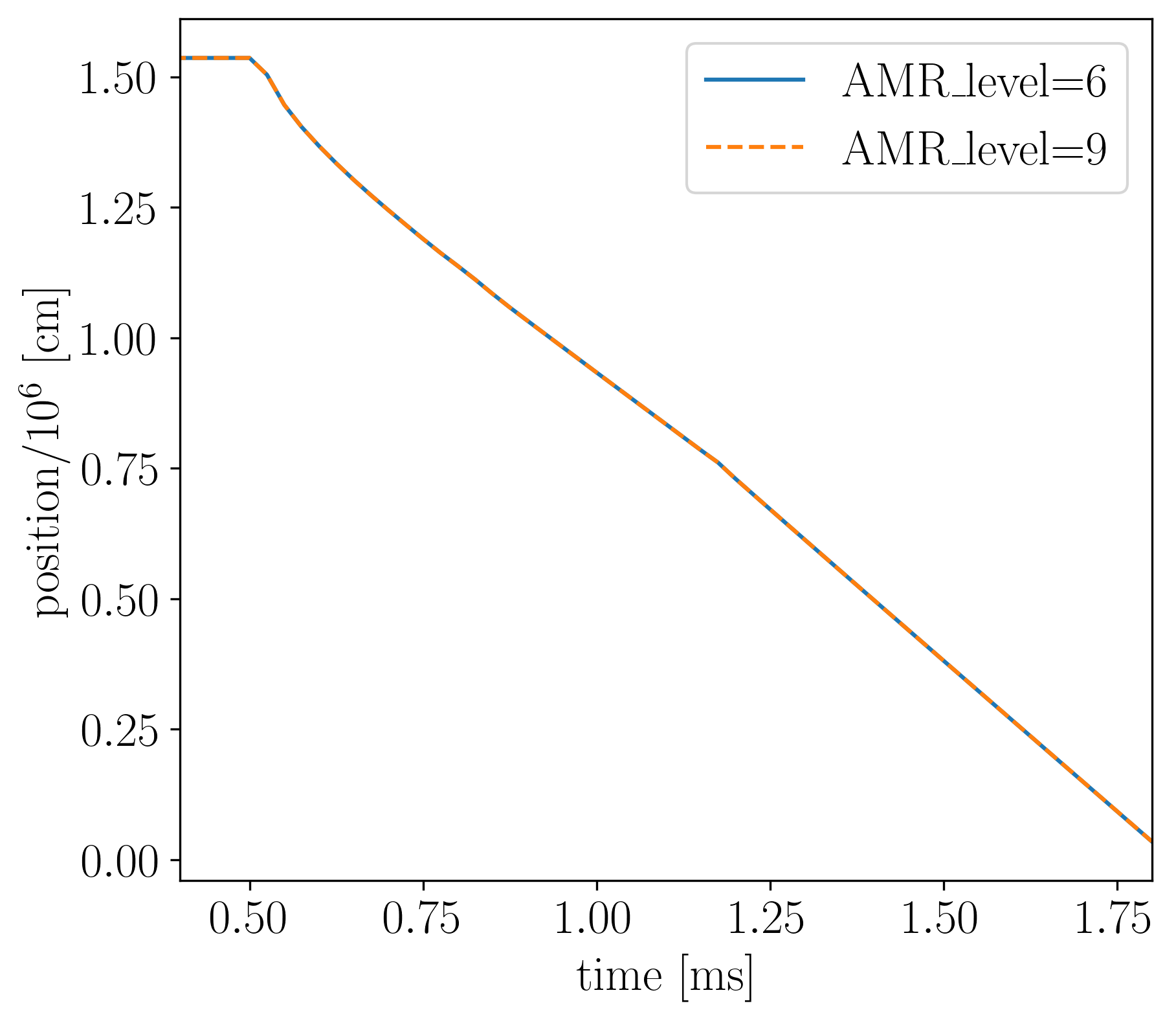}
\caption{Comparison of the oxygen flame front position with AMR level equal to 6 and 9 in one dimension at $\rho_{0}=3.5\times10^{7}\ \rm{g\ cm^{-3}}$.}
\label{Fig.O_pos_1d}
\end{figure}

In the 1D simulations we increased the AMR level from 6 to 9 and compared the time evolution of the oxygen flame front position. As shown in Figure~\ref{Fig.O_pos_1d}, the front trajectories for AMR levels 6 and 9 are nearly identical. This close agreement indicates that AMR level 6 provides sufficient spatial resolution to capture the oxygen flame propagation in our setup, and justifies the use of AMR level 6 for the parameter survey presented in this work while conserving computational resources.

\FloatBarrier






\end{document}